# Determining geopotential difference via relativistic precise point positioning time comparison: A case study using simulated observations

Chenghui Cai[1,4], Wen-Bin Shen[1,2], Ziyu Shen[3], Wei Xu[1], An Ning[1]

[1] Department of Geophysics, School of Geodesy and Geomatics/Key Laboratory of Geospace Environment and Geodesy of Ministry of Education, Wuhan University, Wuhan 430079, China

[2] State Key Laboratory of Information Engineering in Surveying, Mapping and Remote Sensing, Wuhan University, Wuhan 430079, China

[3] Hubei University of Science and Technology, Xianning 437100, China

[4] Guangxi Key Laboratory of Spatial Information and Geomatics, Guilin 541006, China

Corresponding author: Wen-Bin Shen (e-mail:wbshen@sgg.whu.edu.cn).

**Abstract:** According to general relativity theory (GRT), the geopotential difference (GD) can be determined by comparing the change in time difference between precise clocks using the precise point positioning (PPP) time transfer technique, referred to as the relativistic PPP time comparison approach. We focused on high-precision time comparison between two precise clocks for determining the GD using the relativistic PPP time transfer, and conducted simulation experiments to validate the approach. In the experiments, we consider three cases to evaluate the performance of the approach using clocks with different stabilities, namely, the frequency stabilities of the clocks equipped at three selected ground stations are respectively $4.0 \times 10^{-13}/\sqrt{\tau}$ (Case 1), $2.3 \times 10^{-14}/\sqrt{\tau}$ (Case 2), and $2.8 \times 10^{-15}/\sqrt{\tau}$ (Case 3) at time period $\tau$. Conclusions are drawn from the experimental results. First, high-precision clocks can significantly improve the accuracy for PPP time transfer, but the improvement is limited by measurement noises. Compared to Case 1, the long-term stabilities of OPMT-BRUX as well as PTBB-BRUX are improved in Cases 2 and 3. The frequency stabilities of Cases 1–3 are approximately $4.28 \times 10^{-16}$, $4.00 \times 10^{-17}$, and $3.22 \times 10^{-17}$ at 10-day averaging time for OPMT-BRUX, respectively, and for PTBB-BRUX, these values are approximately $3.73 \times 10^{-16}$, $8.17 \times 10^{-17}$, and $4.64 \times 10^{-17}$. Second, the geopotential difference between any two stations can be determined at the decimeter level, with its accuracy being consistent with the stabilities of the time links in Cases 1–3. In Case 3, the determined geopotential differences between OPMT and BRUX deviate from the EIGEN-6C4 model values by -0.64 m$^2$/s$^2$ with an uncertainty of 1.11 m$^2$/s$^2$, whereas the deviation error between PTBB and BRUX is 0.76 m$^2$/s$^2$ with an uncertainty of 1.79 m$^2$/s$^2$. The approach prosed in this study can be also applied to testing GRT.

## Introduction

The geopotential plays a very important role in geodesy and has broad applications in various fields. The classic method of determining geopotential difference is based on leveling with additional gravimetry (Heiskanen and Moritz 1967), which has drawbacks in that it is labor-intensive and geographically limited (Shen et al. 2016). To overcome these drawbacks, Bjerhammar (1985) proposed that geopotential difference could be determined using a clock transportation comparison (CTC) approach (Shen et al. 2009) based on the general relativity theory (GRT) (Einstein 1915). The CTC technique is based on continuously comparing the change in time difference between a fixed clock with a transportable clock (Grotti et al. 2018). The key problem lies in the difficulties associated with transporting high-precision clocks (Kopeikin et al. 2016) because it is very difficult to control ideal operating conditions (temperature and humidity) during the transportation process (Shen et al. 2011).

Alternative approaches of determining the geopotential difference using clocks require connecting two clocks with optical fiber/coaxial cables or satellites to transmit frequency signals or time signals between two stations (Lisdat et al. 2016; Takano et al. 2016; Shen et al. 2017, 2019; Takamoto et al. 2020). Chou et al. (2010b) observed time dilation by comparing two separate optical atomic clocks connected by a 75-m optical fiber, and they detected a change in height of 37 ± 15 cm compared to the actual height of 33 cm. Atomic clocks with a stability of 1.0 × 10$^{-18}$ may be capable of sensing height variations of one centimeter (McGrew et al. 2018). A time transfer simulation experiment was carried out using two atomic clocks with stabilities of 1.0 × 10$^{-18}$ connected by a coaxial cable, and the results of the simulation experiment achieved an accuracy of 0.16 m$^2$/s$^2$ (equivalent to 1.6 cm in height) (Shen and Shen 2015). Although relative accuracies of up to the $10^{-19}$ level can be achieved for comparisons of time and frequency transfer using optical fibers (Grosche et al. 2009; Calonico et al. 2014), sufficient underground fibers are required to connect two arbitrary stations, which limits its applications, for instance over ocean and mountainous areas.

With the rapid development of the global navigation satellite system (GNSS), receiver clock offsets can be estimated as independent parameters for each measurement epoch (Wang and Rothacher 2013), which could have potential applications in time-frequency science, providing a good opportunity to determine the geopotential difference. According to the studies of Shen et al. (1993, 2011, 2017), the geopotential difference between a ground station and a satellite can be calculated using the satellite frequency signal transmission based on the Doppler-canceling technique or tri-frequency combination technique. Unlike time and frequency transfer using fibers on the ground, time and frequency transfer between ground stations and satellites poses much more challenging problems, such as signal propagation delay and frequency shift, orbital error, and Earth rotation. If a time and frequency transfer accuracy of less than 1 ns or higher is desired, the aforementioned errors must be eliminated or significantly diminished. As precise satellite orbits and clock products can be generated in the frame of the international GNSS service (IGS) (Dow et al. 2005), the precise point positioning (PPP) technique is widely applied to compute the time and frequency links with sub-nanosecond accuracy (Zhang et al. 2018; Tu et al. 2018). With its low noise of phase measurements, PPP can provide much higher short-term stability than the GNSS pseudorange-only technique. Petit et al. (2015) demonstrated that the frequency stability of the integer-PPP technique can reach an accuracy of $1.0 \times 10^{-16}$ within a few days.

Considering the advantages of the high-precision of GNSS PPP time transfer technique and rapid development of time and frequency science, for instance at present optical atomic clocks achieve a stability of $4.8 \times 10^{-17}$ at 1 s and $6.6 \times 10^{-19}$ over an hour-long measurement (Oelker et al. 2019), in this study, we propose an approach that uses the PPP technique to directly compute clock offsets between two clocks at two arbitrary positions for the determination of geopotential difference, referred to as the relativistic PPP time comparison approach, and the accuracy of this approach depends not only on the stabilities of clocks, but also the time transfer technique itself.

In this paper, we first introduce the relativistic PPP time comparison approach, including the relationship between the change in accumulated time difference and the geopotential difference between two remote clocks at two stations, and the principle of PPP time transfer technique. The clocks used should be previously calibrated at same site. Then the clocks run freely at different stations of our interests without any artificial adjustment. Based on this approach, the geopotential difference could be determined. To validate the proposed approach, simulation experiments are conducted. We simulated GNSS observations from ground stations equipped with ultra-high precise clocks (say optical clock), and added various measurement errors to the GNSS observation simulation. The change in time difference between two clocks can be estimated from simulated observations using the PPP time transfer technique, and then the geopotential difference between two stations can be determined. Afterward, we described the simulation strategies of GNSS observations, receiver clock offset model, experimental data, and processing strategies. Thereafter, we comprehensively evaluate the performance of this approach in different cases using clocks with different stabilities, and relevant results are presented. Furthermore, we discuss the requirements in this approach and the relationship between the performance of time links and the accuracy of this approach. Finally, we conclude this work in the last section.

## Methodology

In this section, we first introduce the relativistic time comparison approach for determining the geopotential and then briefly describe the process of determining the change in time difference between two separate clocks using the PPP time transfer technique.

### Determination of Geopotential Difference Between Two Ground Stations based on Clocks' Running Rates

According to GRT, a precise clock runs faster at a position with higher geopotential than an identical clock at a position with lower geopotential (Shen et al. 2009). Suppose there are two clocks $C_P$ and $C_Q$ at two different stations P and Q, respectively, and one clock $C_0$ on the geoid (an equi-geopotential surface nearest to the mean sea level) on which the geopotential constant is noted as $W_0$, and the value of $W_0$ is 62636851.71 m²/s² (Petit and Luzum 2010; Sánchez et al. 2014). After a standard time duration $dt_0$, recorded by clock $C_0$, clocks $C_P$ and $C_Q$ will have time durations $dt_P$ and $dt_Q$, respectively. Accurate to the level of $c^{-2}$, $dt_P$ and $dt_Q$ could be expressed respectively as (Vessot and Levine 1979; Bjerhammar 1985):

$$dt_P = \left(1 + \frac{W_{0P}}{c^2}\right) dt_0 \tag{1}$$

$$dt_Q = \left(1 + \frac{W_{0Q}}{c^2}\right) dt_0 \tag{2}$$

where $W_{0P} = W_0 - W_P$ and $W_{0Q} = W_0 - W_Q$ represent the geopotential numbers between the geoid and the stations P and Q, respectively; $W_P$ and $W_Q$ denote the geopotentials at stations P and Q; and $c$ is the speed of light in vacuum. In this study, the definition of the geopotential in physical geodesy is applied: it always holds that $W \geq 0$, which is different from the definition in physics (Shen et al. 2011). Thus, if station P (or Q) is above the geoid, it holds that $W_P < W_0$ (or $W_Q < W_0$) (Pavlis and Weiss 2003). Based on Eqs. (1–2), after standard time duration $T$, the clock offset between clocks $C_P$ and $C_Q$ can be described as:

$$\begin{aligned} \Delta t_{PQ} &= t_Q - t_P \\ &= \int_0^T \left(1 + \frac{W_{0Q}}{c^2}\right) dt_0 - \int_0^T \left(1 + \frac{W_{0P}}{c^2}\right) dt_0 \\ &= -\frac{\Delta W_{PQ}}{c^2} T \end{aligned} \tag{3}$$

where $t_P$ and $t_Q$ are the clock offset of clocks $C_P$ and $C_Q$, respectively. Then, the geopotential difference between P and Q can be formulated as:

$$\Delta W_{PQ} = -\frac{\Delta t_{PQ}}{T} c^2 \tag{4}$$

where $\Delta W_{PQ} = W_Q - W_P$. Hence, if GRT holds, we can determine the geopotential difference between the two stations by comparing the change in time difference $\Delta t_{PQ}$ after standard time duration $T$ using precise clocks.

Suppose the orthometric height of P (noted as $H_P$) is given, the orthometric height of Q (noted as $H_Q$) can be determined, expressed as:

$$\begin{cases} H_P = -\dfrac{W_P - W_0}{\bar{g}_P} \\ H_Q = -\dfrac{W_Q - W_0}{\bar{g}_Q} = -\dfrac{\Delta W_{PQ} - (W_0 - W_P)}{\bar{g}_Q} \end{cases} \quad (5)$$

where $\bar{g}_P$ and $\bar{g}_Q$ are the mean value of the gravity along the plumb line, respectively. If $\bar{g}_i (i = P, Q)$ can be determined, then the $H_i$ can be computed. The mean gravity $\bar{g}_i$ can be formulated as:

$$\bar{g}_i = \dfrac{1}{H_i} \int_0^{H_i} g(h) \mathrm{d}h \quad (6)$$

where $g(h)$ is the actual gravity at the station $i$ which has the height $h$. Since we cannot determine the mean gravity $\bar{g}_i$ precisely, Eq. (6) can be approximated as (Hofmann-Wellenhof and Moritz 2005):

$$\bar{g}_i = g_i + 0.0424 H_i \quad (7)$$

where $g_i$, in gal, is the gravity measured at the ground station $i$, which can be measured by absolute gravimeter, and $H_i$ in km. The factor 0.0424 refers to the normal density $\rho = 2.67 \ g/cm^3$. According to Eqs. (5) and (7), we obtain the orthometric height $H_Q$, then the practically useful formula can be expressed as:

$$H_Q = -\dfrac{\Delta W_{PQ} - H_P(g_P + 0.0424 H_P)}{g_Q + 0.0424 H_Q} \quad (8)$$

where $\Delta W_{PQ}$ is measured in geopotential units (g.p.u.), and 1 g.p.u.=1000 gal.m. We can observe that the accuracy of the determined $\Delta W_{PQ}$ will affect that of $H_Q$, then the determination of geopotential difference has the potential applications in orthometric height determination and in unifying the world vertical height system (WVHS).

The key problem is to determine the change in time difference between two remote clocks using the GNSS satellites microwave signal. In order to eliminate the errors involved in GNSS time transfer, here we use the GPS ionosphere-free combination PPP model (IF-PPP).

**GPS Ionosphere-free PPP model**

The undifferenced (UD) observation equations of the dual-frequency pseudorange $P_{r,i}^s$ and carrier phase $L_{r,i}^s$ measurements at one epoch can be respectively modeled as (Leick et al. 2015):

$$P_{r,i}^s = \rho_r^s + c \cdot (dt_r - dt^s) + T_r^s + \gamma_i^s \cdot I_{r,1}^s + c \cdot (d_{r,i} - d_i^s) + \varepsilon_{r,i}^s \tag{9}$$

$$L_{r,i}^s = \rho_r^s + c \cdot (dt_r - dt^s) + T_r^s - \gamma_i^s \cdot I_{r,1}^s + \lambda_i \cdot (N_{r,i}^s + b_{r,i} - b_i^s) + \xi_{r,i}^s \tag{10}$$

where indices $s$, $r$, and $i$ denote the satellite, receiver, and carrier frequency, respectively; $P_{r,i}^s$ and $L_{r,i}^s$ denote the pseudorange and carrier phase measurements, respectively, in meters, which need to be simulated at each epoch; $\rho_r^s$ represents the geometric distance between the phase centers of the satellite $s$ and receiver $r$ antennas; $c$ is the speed of light in vacuum; $dt_r$ and $dt^s$ denote respectively the clock offsets between the receiver clock and GPS time (GPST) and the satellite clock and GPST; $T_r^s$ is the tropospheric delay of the signal path in meters; $I_{r,1}^s$ is the slant ionospheric delay on frequency $f_1^s$; $\gamma_i^s$ is the ionospheric factor depending on the frequency $f_i$, $\gamma_i^s = (f_1^s/f_i^s)^2$; $\lambda_i$ is the carrier phase wavelength at frequency $f_i$; $N_{r,i}^s$ is the integer phase ambiguity in cycles; $d_{r,i}$ and $d_i^s$ are the code delays of receiver and satellite, respectively, in meters; $b_{r,i}$ and $b_i^s$ are the uncalibrated phase delays (UPDs) at receiver and satellite, respectively in cycles; and $\varepsilon_{r,i}^s$ $\xi_{r,i}^s$ refer to the multipath effects and unmodeled measurement errors for pseudorange and carrier phase observations, respectively in meters. In addition, some errors, such as the dry component of tropospheric delays, tidal loading, phase wind-up, relativistic effects in the satellite clock, phase center offsets (PCOs), and phase center variations (PCVs), are not included in Eqs. (9–10). These errors can be also corrected by relevant models (Kouba 2009; Petit and Luzum 2010).

The IF-PPP model is most generally used for time and frequency transfer, which can eliminate first-order ionospheric delays (Kouba and Héroux 2001), expressed as:

$$\begin{cases} P_{r,IF}^s = \alpha \cdot P_{r,1}^s + \beta \cdot P_{r,2}^s \\ L_{r,IF}^s = \alpha \cdot P_{r,1}^s + \beta \cdot P_{r,2}^s \\ d\tilde{t}_r = dt_r + \alpha \cdot dt_{r,1} + \beta \cdot dt_{r,2} \\ d\tilde{t}^s = dt^s + \alpha \cdot d_1^s + \beta \cdot d_2^s \end{cases} \quad (11)$$

where $\alpha$ and $\beta$ are ionosphere-free combination coefficients with $\alpha = f_1^2/(f_1^2 - f_2^2)$ and $\beta = -f_2^2/(f_1^2 - f_2^2)$; $P_{r,IF}^s$ and $L_{r,IF}^s$ refer to ionosphere-free pseudorange and carrier phase observations, respectively, in meters; and $d\tilde{t}_r$, $d\tilde{t}^s$ are respectively receiver and satellite clock offsets, which absorb ionosphere-free code hardware delays.

**Determination of Geopotential Difference Between Two Ground Stations via GNSS Satellites**

As shown in Fig. 1, there are two ground stations P and Q, which can receive the signals emitted from GNSS satellites. To determine the geopotential difference between two stations using the relativistic PPP time comparison approach, we performed the following procedures.

(i) The freely running clocks $C_P$ and $C_Q$ are used at stations P and Q, respectively; the clocks should be a priori synchronized at same site (say at station P) before the experiment, and their vibration frequencies should not be adjusted during the whole experiment. Hence, the clock's stability is significant in this study. Its accuracy does not matter.

(ii) After synchroniation at station P, the clock $C_P$ is fixed at P and clock $C_Q$ is transported to station Q, both clocks outputing local 1 pulse per second (1 PPS) signals, and the GNSS receivers linked with precise clocks obtain the observations from GNSS satellites signals.

(iii) We employed an open-source program RTKLIB (Takasu and Yasuda 2009) which can perform GNSS IF-PPP processing to compute the time offsets between the clock $C_P$ (or $C_Q$) at a ground station and the GPST, where the GPST is taken as reference. By subtracting one time offset series from another we obtain the change in time difference between the remote clocks $C_P$ and $C_Q$ on ground, cancelling the "common" GPST. Here we see that the

GPST is used only as reference (or "bridge"), having no effect on the time offsets between two remote clocks.

(iv) The corresponding geopotential difference between stations P and Q is determined based on Eq. (4).

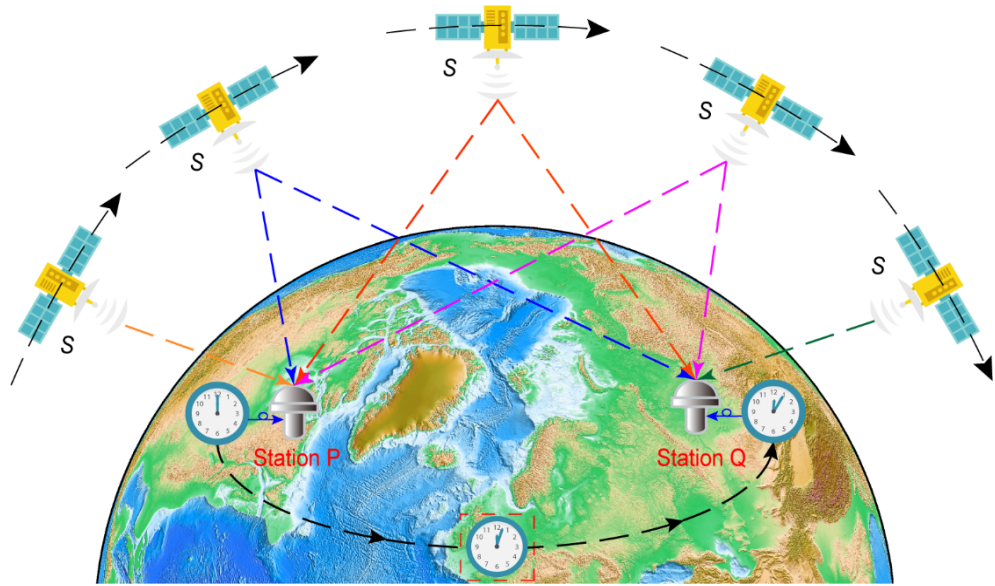

**Fig.1** Principle of determining geopotential difference between stations P and Q via GNSS satellite signal transmission. Two clocks are synchronized at station P, and one clock is transported to station Q after synchronization. Both clocks outputting local 1 PPS signals, and receivers receiving the signals from GNSS satellites to extract the observations. The background is based on the earth relief dataset provided with Generic Mapping Tools (Wessel et al. 2013).

## The Strategy of Data Simulation

Although some ground stations are equipped with high-precision clocks, their clocks are frequently adjusted to conform with the Coordinated Universal Time (UTC). Hence, the observations at present IGS stations cannot be used to determine the geopotential. In another aspect, at present the stability of the GNSS PPP frequency transfer is limited to about $1.0 \times 10^{-15}$ at 1 day averaging time (Petit et al. 2015), which is insufficient to validate the relativistic PPP time comparison approach at the decimeter level. These difficulties could be overcome by using free-running clocks with ultra-high stability (say $1.0 \times 10^{-18}$ /d). Here in this study, we simulated GNSS observations from ground stations equipped with ultra-high-precision clocks (e.g., optical clocks). GNSS observations associated with ground station clocks were simulated to evaluate the performance of this approach, and the flowchart of determining the geopotential difference between two ground stations using this approach is shown in Fig. 2.

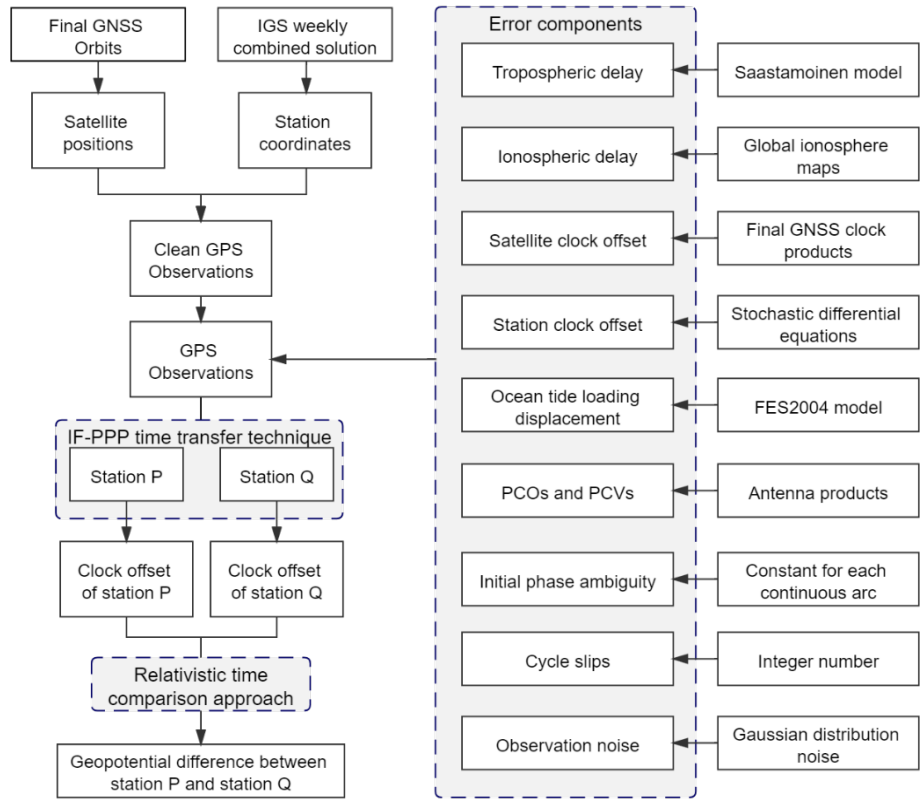

**Fig.2** Principle of the relativistic PPP time comparison approach, including the GNSS observation simulation processing and geopotential difference determination between two stations, P and Q.

**Simulation of GNSS Observations Connected with Ground Stations Clocks**

In this subsection, we describe the simulation of GNSS observations, which is essentially the inverse process of positioning. As mentioned above, all measurement errors, including satellite-dependent, receiver-dependent, and atmospheric propagation errors, were corrected according to existing models (Kouba 2009). GNSS observations were used to estimate the receiver position and clock offset in the GNSS positioning process. For simulating GNSS observations, if the receiver/satellite position and clock are known values, the satellite-receiver geometric distance can be computed. To simulate the observations as accurately as possible, various measurement errors were calculated using existing models and added to the geometric distance with random noise to generate the observations (Li et al. 2019).

In the observation simulation process, we focused on simulating all components on the right side of Eqs. (9–10). The satellite-receiver geometric range is calculated by the positions of the GNSS satellite and ground receiver. The positions of the GNSS satellite can be extracted by final GNSS orbit products with a sampling rate of 5 min, acquired from the Center for Orbit Determination in Europe (CODE). Nevertheless, satellite positions need to be interpolated to access the satellite positions at any epoch. In addition, receiver positions can be obtained from weekly solutions of IGS stations. Tropospheric delay is composed of dry and wet components, both of which can be formulated as zenith delay and the corresponding mapping function. In terms of zenith dry delay, which can be calculated by combining the global pressure and temperature (GPT) empirical model with the Global Mapping Function (GMF) (Boehm et al. 2006, 2007), according to the Saastamoinen model (Saastamoinen 1972). Zenith wet delay can be obtained from GNSS PPP solutions. Ionospheric delay is associated with the total electron content (TEC), which can obtain from the global ionosphere maps (GIM) of CODE. The initial ambiguity of phase measurement is set to an integer number of cycle constant for each continuous arc, and cycle slips are also introduced at some epochs for certain satellites. The satellite clock offsets are obtained from CODE precise clock products, and the receiver clock offsets are simulated by stochastic differential equations (SDEs) (Zucca and Tavella 2015), which are described in the next subsection. Moreover, some errors, such as the ocean tide loading displacement, phase center offsets (PCOs), and phase center variations (PCVs), need to be included in the observation errors to ensure that the simulated observation data are as close to real data as possible. The ocean tide loading displacement can be calculated by the FES2004 ocean tide model (Lyard et al. 2006). Antenna products from IGS are employed in PCOs and PCVs simulations for satellites and receivers (Petit and Luzum 2010). To generate realistic data, the observation noises are simulated as a zero-mean-value random measure noises with Gaussian distribution, for which the standard deviation (STD) depends on satellite elevation angle. Specifically, STD decreases with increasing elevation angle. In the zenith direction, the STD of each frequency is set to 0.2 m and 2 mm for code and phase observations, respectively.

**Mathematical Model of Clock Offset**

In this subsection, we introduce the mathematical model of the clock offset, which is very useful for clock simulation. Clock noises are typically affected by five random noises (Allan 1987) generally known as white phase modulation (WPM), flicker phase modulation (FPM), white frequency modulation (WFM), flicker frequency modulation (FFM), and random walk frequency modulation (RWFM). These random noises vary according to the type of clock. Previous studies have reported that WFM and RWFM are the predominant noises in high-precise atomic clocks (Zucca and Tavella 2005, 2015). Accordingly, the mathematical model of the clock offset can be formulated as (Zucca and Tavella 2015):

$$\begin{cases} \mathrm{d}X_1(t) = (X_2(t) + \mu_1)\mathrm{d}t + \sigma_1 \mathrm{d}W_1(t) \\ \mathrm{d}X_2(t) = \mu_2 \mathrm{d}t + \sigma_2 \mathrm{d}W_2(t) \end{cases}, \quad t \geq 0 \qquad (12)$$

with initial conditions

$$\begin{cases} X_1(0) = c_1 \\ X_2(0) = c_2 \end{cases}$$

where $X_1(t)$ represents the clock phase deviation, and $X_2(t)$ is a part of the clock frequency deviation, which is generally called the random walk component; the constants $\mu_1$ and $\mu_2$ can be interpreted as drift terms for the two Wiener processes, referred to as the deterministic component driving clock errors in general. In particular, $\mu_1$ is related to the constant initial frequency offset, indicated by $y_0$; $\mu_2$ is generally indicated by $d$, and it is the frequency drift, also termed as aging. In more familiar metrological notation, $c_1 + \mu_1 = y_0$ and $\mu_2 = d$. $W_1(t)$ is the Wiener process on the clock phase deviation $X_1(t)$ driven by the white noise of frequency and corresponds to WFM, whereas $W_2(t)$ denotes the Wiener noise of frequency corresponding to RWFM, which produces an integrated Wiener process on the phase. The constants $\sigma_1$ and $\sigma_2$ are the diffusion coefficients of the two noise components $W_1$ and $W_2$, respectively, and indicate the intensity of WFM and RWFM. We note that clock offset is dominantly affected by WFM, RWFM, and the deterministic trend terms $\mu_1$ and $\mu_2$.

The relationship between the Allan deviation (ADEV) and the diffusion coefficients of the SDEs can be expressed as (Zucca and Tavella 2005):

$$\begin{cases} \sigma_y^{WFM}(\tau) = \sqrt{\dfrac{\sigma_1^2}{\tau}} \\ \sigma_y^{RWFM}(\tau) = \sqrt{\dfrac{\tau \sigma_2^2}{3}} \end{cases} \quad (13)$$

where $\sigma_y^{WFM}(\tau)$ and $\sigma_y^{RWFM}(\tau)$ denote the ADEVs related to the WFM and RWFM at averaging time $\tau$, respectively.

To evaluate the performances of the relativistic PPP time comparison approach for different receiver clocks with different stabilities, we propose three cases as summarized in Table 1. The deterministic component $\mu_1$ is considered as the gravitational frequency shift $\Delta t/T$, which can be solved by Eq. (4). As the value of frequency drift $d$ does not affect the uncertainty of the frequency drift (Zucca and Tavella 2005; Wu et al. 2015), we can simply assume that $d=0$. We simulated the clocks in Case 1 with WFM having a fractional frequency stability of $\sigma_y^{WFM}(\tau) = 4.0 \times 10^{-13}/\sqrt{\tau}$ and RWFM of $\sigma_y^{RWFM}(\tau) = 4.0 \times 10^{-19}\sqrt{\tau}$, both with averaging time $\tau$, which is typical Cs-fountain performance (Yao et al. 2018). The clocks in Cases 2 and 3 were simulated with WFM having fractional frequency stabilities of $\sigma_y^{WFM}(\tau) = 2.3 \times 10^{-14}/\sqrt{\tau}$ and $\sigma_y^{WFM}(\tau) = 2.8 \times 10^{-15}/\sqrt{\tau}$, respectively, and RWFMs of $2.3 \times 10^{-20}\sqrt{\tau}$ and $2.8 \times 10^{-21}\sqrt{\tau}$ at averaging time $\tau$, which are typical for optical clocks (Chou et al. 2010a; Cao et al. 2017). It should be noted that the simulation parameters of clocks in Case 1–3 are applied for whole experiment period, without artificially steering outputs of clocks to approximate UTC.

**Table 1** Three cases of clock fractional frequency stability at averaging time $\tau$ for simulation experiments

| Experiment | WFM | RWFM |
|---|---|---|
| Case 1 | $4.0 \times 10^{-13}/\sqrt{\tau}$ | $4.0 \times 10^{-19}\sqrt{\tau}$ |
| Case 2 | $2.3 \times 10^{-14}/\sqrt{\tau}$ | $2.3 \times 10^{-20}\sqrt{\tau}$ |
| Case 3 | $2.8 \times 10^{-15}/\sqrt{\tau}$ | $2.8 \times 10^{-21}\sqrt{\tau}$ |

## Experimental Data and Processing Strategies

To evaluate the performance of the PPP time transfer technique and validate the tests of geopotential determination via relativistic PPP time comparison approach, we selected three IGS stations that can provide fixed station coordinates from IGS weekly solutions for ground-based observation simulation. The geographical distribution of the selected stations is shown in Fig. 3, and these stations provide GNSS data with a sampling rate of 5 s. Table 2 summarizes the details of the selected IGS stations, including the latitude, longitude, height (here it denotes the height above geoid), and geopotential ($W$). The geopotentials at all selected stations were calculated using the new global combined gravity field model EIGEN-6C4. The accuracy of the EIGEN-6C4 model is approximately 10–20 cm in land areas (Förste et al. 2014), which is sufficient for the precision requirement of this study. The experimental data cover a 30–day period for the day of year (DOY) 061–090 in 2020 (during Modified Julian Date (MJD) 58909–58938).

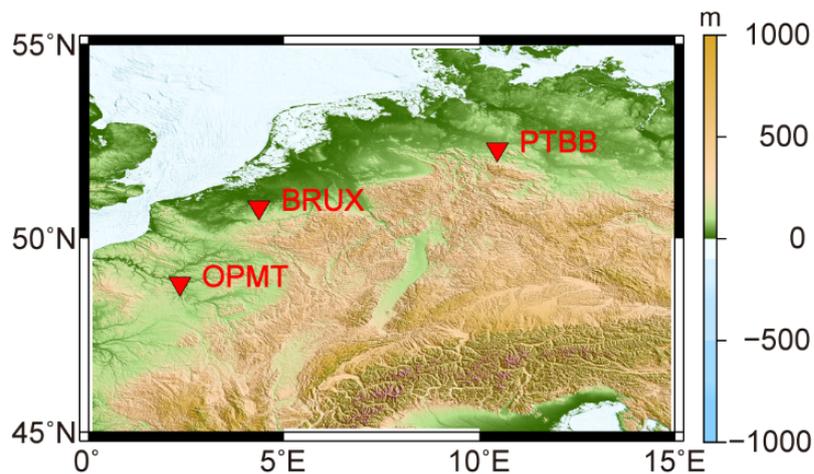

**Fig. 3** Geographical distribution of three selected IGS GNSS stations. BRUX at Royal Observatory of Belgium (ROB), OPMT at Observatoire de Paris (OP), and PTBB at Physikalisch-Technische Bundesanstalt (PTB)

**Table 2** Information of stations

| Station | Lat (deg) | Long (deg) | Height (m) | $W$ (m²/s²) |
|---------|-----------|------------|------------|-------------|
| BRUX | 50.798 | 4.359 | 158.100 | 62635750.739 |
| OPMT | 48.836 | 2.335 | 122.600 | 62636090.521 |
| PTBB | 52.296 | 10.460 | 130.200 | 62636000.310 |

In addition, the detailed PPP processing strategy is summarized in Table 3, including the estimated parameters and observation models. Using the simulated GPS observations, we can analyze performances of both the PPP time transfer technique and geopotential difference determination with different receiver clock cases. In the data processing, GPS PPP solutions were based on dual-frequency (L1/L2) ionosphere-free combinations for eliminating first-order ionospheric delay. The initial standard deviations for the raw GPS carrier phase and code observations were 2 mm and 0.2 m, respectively. Precise satellite orbit and clock products are provided by CODE at sampling rates of 5 min and 5 s, respectively.

**Table 3** Detailed PPP processing strategy

| Items | Models |
|-------|--------|
| Solution | Ionosphere-free PPP |
| Observations | Pseudorange and carrier phase observations |
| Sampling rate | 5 s |
| Elevation cutoff | $7°$ |
| Weighting scheme | Elevation-dependent weight |
| Ionospheric delay | First-order ionospheric delay eliminated by Ionosphere-free combination (Kouba and Héroux 2001) |
| Tropospheric delay | Dry component: corrected with Saastamoinen model (Saastamoinen 1972) Wet component: estimated as random-walk process, GMF mapping function applied |
| Satellite antenna error | Correct with conventional PCO and PCV from IGS14.atx |
| Receiver antenna error | Correct with conventional PCO and PCV from IGS14.atx |
| Receiver clock offset | Estimate as white noise process |
| Tide | Corrected (Lyard et al. 2006) |
| Sagnac effect | Corrected (Petit and Luzum 2010) |

| Phase ambiguities | Estimated as constants; float value |
| Station coordinate | Estimated as constants for each arc |

## Validation and Analysis

We first evaluated the performances of the clocks in different cases. With the simulated GNSS observation data, we analyzed the performance of the PPP time transfer technique with different cases, and the results of the relativistic PPP time comparison approach are presented.

### Performance Evaluation of Clock Offsets Simulation

In order to analyze the performances of the relativistic PPP time comparison approach with different receiver clocks, GNSS observations need to be simulated with different receiver clock offsets, as summarized in Table 2. Based on the SDEs, we simulated the receiver clock offsets over a 30-day period and evaluated the performance. The frequency stabilities of the simulated clock offsets, shown in Fig. 4, included total Allan deviation (TADEV), along with WFM asymptotes for Cases 1–3. The WFM asymptotes for Cases 1–3 follow $4.0 \times 10^{-13}/\sqrt{\tau}$, $2.3 \times 10^{-14}/\sqrt{\tau}$, and $2.8 \times 10^{-15}/\sqrt{\tau}$ at every averaging time $\tau$, respectively, which are consistent with simulation parameters of WFM.

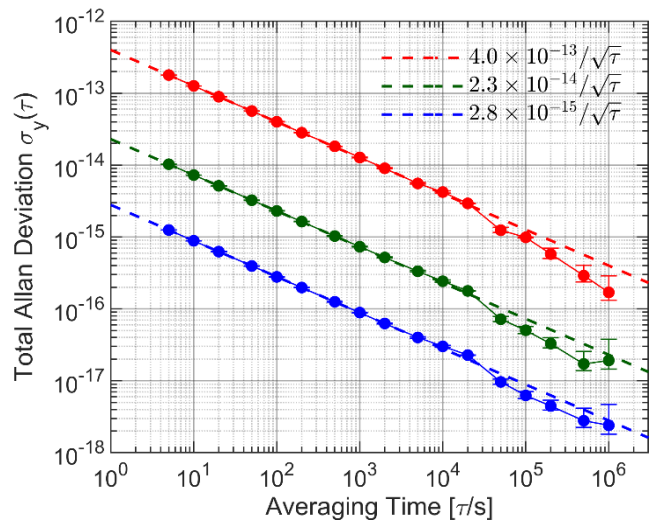

**Fig. 4** Frequency stabilities of simulated clock offsets in three cases calculated according to TADEV. Red, green, and blue circles represent the TADEV of Cases 1–3, respectively. Red,

green, and blue dashed line are the WFM asymptotes for Cases 1–3, respectively, where $\tau$ is the averaging time in seconds. Error bars represent the 1-σ uncertainty in TADEV

To further evaluate the performances of three cases for simulating clock offset, we simulated 500 independent clock offsets for each case based on the SDEs. Fig. 5 presents the 500 clock offsets, and the 1-σ uncertainty (68% confidence level) and 2-σ uncertainty (95% confidence level) of the 500 clock offsets for a 30-day period. For this period, the 1-σ uncertainties of Cases 1–3 were approximately 646.25 ps, 36.97 ps, and 4.44 ps, respectively, which correspond to geopotential difference determination errors of ±22.41 $m^2/s^2$, ±1.28 $m^2/s^2$, and ±0.15 $m^2/s^2$. Therefore, considering Figs. 4–5, it can be concluded that the instability of the clocks decreases with increasing averaging time, and the uncertainty of clock offsets increases with experiment time. Nevertheless, both the instability and uncertainty of the clocks can be improved by increasing the stability of the clocks.

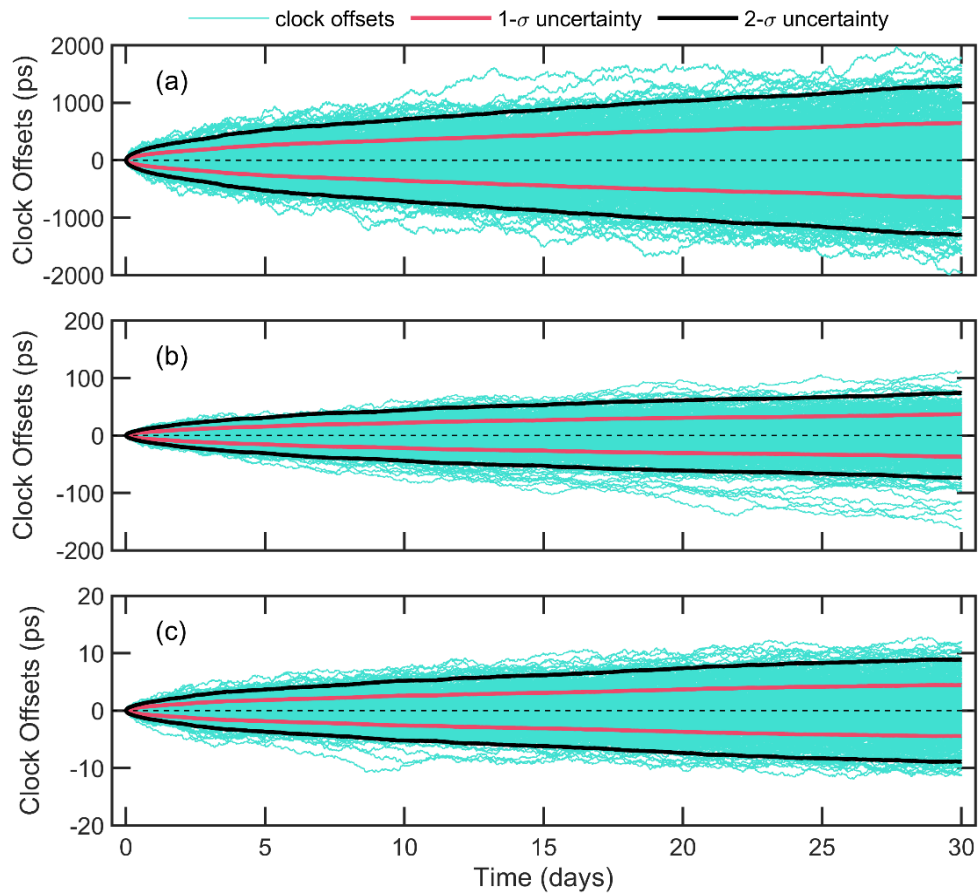

**Fig. 5** 1-σ uncertainty (red curve) and 2-σ uncertainty (black curve) of 500 clock offsets. Subfigures (a)–(c) represent Cases 1–3, respectively

**Accuracy Assessment of PPP Time Transfer in Different Cases**

We evaluated the performance of PPP time transfer in different cases against true values. True values mean that $\sigma_y^{WFM}$ and $\sigma_y^{RWFM}$ are set to 0. The performance of BRUX station in Cases 1–3 between MJD 58909–58938 is presented in Fig. 6. Fig. 6 (a) shows that the clock offset deviations of Case 1 are greater than those of Cases 2 and 3. Moreover, because the accuracy of clock offset is affected by measurement noise, the clock offset deviations of Case 2 are consistent with those of Case 3. Deviations in Case 1 mostly ranged from -0.5 ns to 0.1 ns, but those in Cases 2 and 3 remained within approximately 0.1 ns for the entire period. Fig. 6 (b) presents the frequency stability of BRUX in Cases 1–3. We observed that only the short-term stabilities (within averaging of 1000 s) of Cases 1–3 were very close to each other; the stabilities of Cases 2 and 3 were higher than that of Case 1 for averaging time at and above 1000 s.

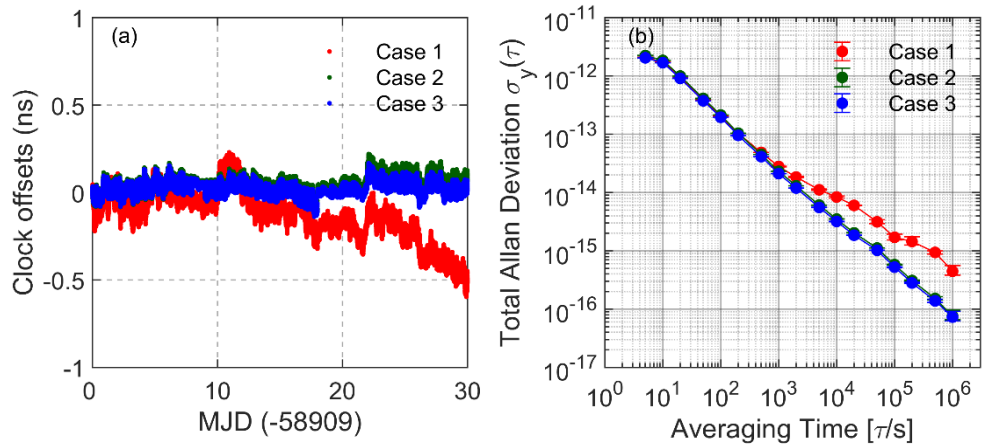

**Fig. 6** Performance of BRUX station in different cases between MJD 58909–58938. (a) Clock offsets obtained from Cases 1–3 with respect to the true values; (b) results of frequency stability analysis of BRUX in Cases 1–3, expressed as TADEV

To further evaluate the performance of the three cases, we calculated the STD and RMS of clock offsets of Cases 1–3 with respect to the true values for the three stations, and the results are presented in Fig. 7. The STD of the clock offsets obtained from Case 1 ranged from 0.1 ns to 0.3 ns, for the three stations, whereas the RMS ranged from 0.2 ns to 0.5 ns. The performance of Case 3 was the best among the three cases, but the STD and RMS of Cases 2 and 3 were very similar for the three stations. The STDs of both Cases 2 and 3 were within approximately 50 ps and the RMS ranged from 40 ps to 100 ps.

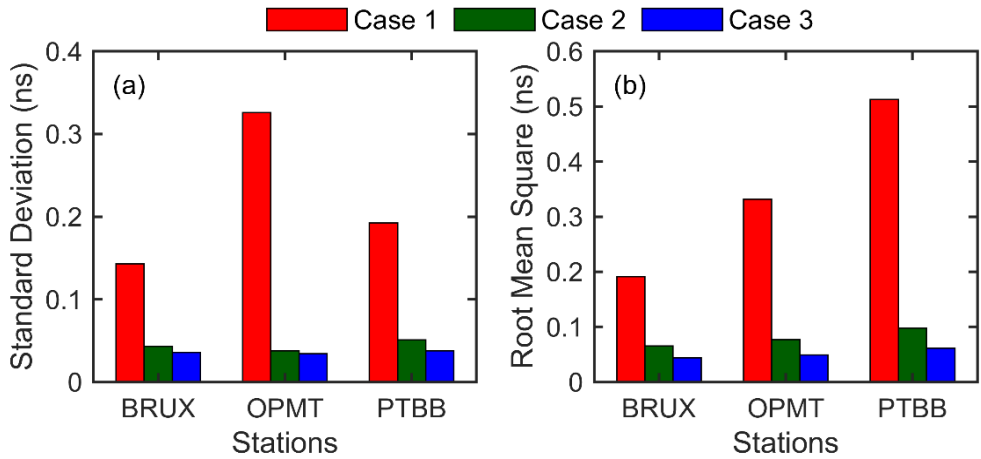

**Fig. 7** STD and RMS of clock offsets of Cases 1–3 with respect to the true values at three stations (BRUX, OPMT, and PTBB)

We used Cases 1–3 to respectively compute the clock offsets of two time transfer links, for which BRUX was selected as the reference clock. The performance of time links in Cases 1–3 was evaluated. Fig. 8 shows that the clock offsets of time links obtained from Cases 1–3 with respect to the true values, and Fig. 9 shows the frequency stability of two time links. In addition, Table 4 summarizes the STD and RMS values of clock offsets for each link. Overall, three conclusions can be drawn from the results. First, the clock offsets of both Cases 2 and 3 are fairly consistent with the true values, and the clock offset deviations of Cases 2 and 3 varied from -0.1 ns to 0.1 ns, for both OPMT–BRUX and PTBB–BRUX. In contrast, the deviations of Case 1 of OPMT–BRUX and PTBB–BRUX ranged from -0.5 ns to 0.6 ns and from -0.1 ns to 1 ns, respectively. Second, the frequency stabilities of Cases 2 and 3 for each link were higher than those for Case 1 at averaging time of 1000 s and above. The frequency stabilities of Cases 1–3 were approximately $4.28 \times 10^{-16}$, $4.00 \times 10^{-17}$, and $3.22 \times 10^{-17}$, respectively, at 10-day averaging for OPMT–BRUX. For PTBB–BRUX, these values were approximately $3.73 \times 10^{-16}$, $8.17 \times 10^{-17}$, and $4.64 \times 10^{-17}$. According to previous studies (Bjerhammar 1985; McGrew et al. 2018), a clock with a stability of $1.0 \times 10^{-16}$ may sense one-meter height variations, according to GRT. The deviation errors of geopotential difference were approximately 38.50 $m^2/s^2$, 3.60 $m^2/s^2$, and 2.90 $m^2/s^2$ for OPMT–BRUX in Cases 1–3, respectively, after a 10-day test period; for PTBB–BRUX, the errors were approximately 33.56 $m^2/s^2$, 7.34 $m^2/s^2$, and 4.17 $m^2/s^2$. Third, the maximum values of STD and RMS for each link were observed in Case 1, and the minimum values were observed in Case 3. Therefore, we can conclude that the clock offsets of stations and links obtained from Case 3 are more stable than those obtained from Case 1, Cases 2 and 3 can provide much higher long-term stability owing to the higher precision clocks, and the frequency stabilities of Cases 2 and 3 facilitate the determination of geopotential difference at the decimeter-level after 10-days of observations.

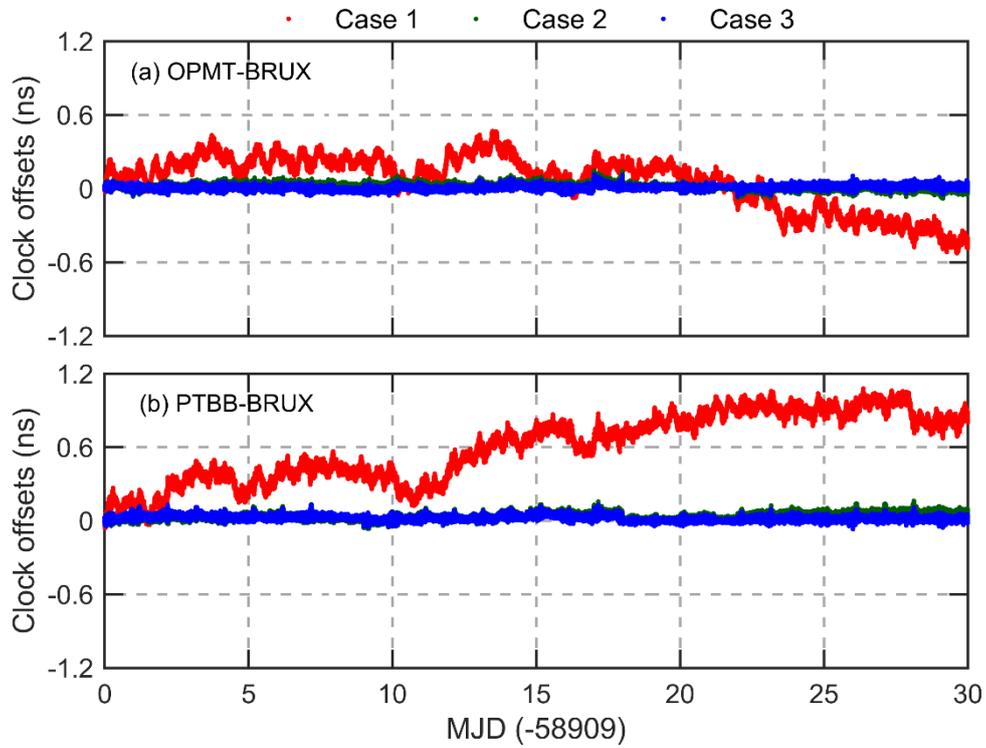

**Fig. 8** Clock offsets of two time transfer links (OPMT–BRUX and PTBB–BRUX) obtained from Cases 1–3 in MJD 58909–58938

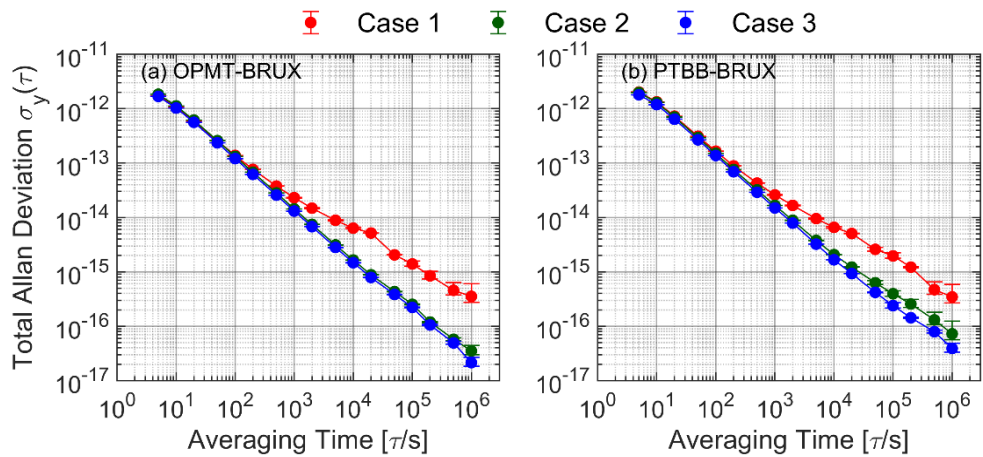

**Fig. 9** Frequency stability analysis of OPMT–BRUX and PTBB–BRUX time links for Cases 1–3 calculated according to TADEV

**Table 4** STD and RMS of clock offsets of Cases 1–3 with respect to the true values at two links (ps)

| Links | STD | | | RMS | | |
|---|---|---|---|---|---|---|
| | Case 1 | Case 2 | Case 3 | Case 1 | Case 2 | Case 3 |
| OPMT-BRUX | 213.90 | 17.93 | 13.61 | 223.16 | 25.45 | 16.92 |
| PTBB-BRUX | 273.36 | 21.81 | 17.48 | 660.81 | 41.10 | 28.74 |

**Validation of Relativistic PPP Time Comparison Approach**

To determine the geopotential difference between two stations based on GRT, the change in time difference between the clocks at two stations during the test period need to be calculated. For this purpose, we obtained clock offsets via PPP time transfer with Cases 1–3 and analyzed performance of clock offsets. We conducted a geopotential difference experiment to test the validity of the relativistic PPP time comparison approach. From the beginning of the experiment, we measured the geopotential difference every day to evaluate the performance of the experimental results with increasing time. The results of the determined geopotential difference deviating from those of the EIGEN-6C4 model are shown in Fig. 10 and Tables 5–6. Error bars represent the 1-σ uncertainty obtained from the TADEV of time links, as shown in Fig. 9. We performed an experiment using 30-day observations, and TADEV was calculated for about one third of the whole period, namely 10-day period at an average time of $10^6$ s. Frequency stability after the averaging time of $10^6$ s can be obtained by fitting the values of TADEV using the least squares method, and extrapolating the fit to the entire experimental period (McGrew et al. 2018). As shown in Fig. 10, we observed that the deviation errors of geopotential differences between OPMT and BRUX and between PTBB and BRUX decrease with increasing time in all cases. In general, the largest errors of geopotential differences as well as the largest uncertainties were obtained at the beginning of the experiment because of errors in the observation data and insufficient data for estimation. In addition, the deviation errors became stable and close to zero for Cases 2 and 3 after ten days of experimental time, but the errors for Case 1 continued to fluctuate due to the limited clock accuracy and measurement noise. Tables 5–6 present the results of the experiment for 1-day, 5-day, 10-day, 20-day, and 30-day test periods. From these results, three conclusions can be directly drawn. First, because each ground station is equipped with a clock with the same noise level and the same parameters were applied for simulating observation data, the deviation errors and uncertainty of each link are similar in each case. Second, the deviation error decrease as clock precision increases. Case 3 shows the highest performance, whereas Case 1 has the largest deviation error as well as uncertainty. In Case 3, the deviation error of measured geopotential differences between OPMT and BRUX was -0.64 $m^2/s^2$ with an uncertainty of 1.11 $m^2/s^2$, and the deviation error between PTBB and BRUX was 0.76 $m^2/s^2$ with an uncertainty of 1.79 $m^2/s^2$. It is noteworthy that although the deviation error using the 20-day observations is smaller than that using the 30-day observations, the uncertainty of the former is higher (see Tables 5 and 6). Third, the accuracy of geopotential difference achieved using the relativistic PPP time comparison approach is consistent with the stabilities of the time links in Cases 1–3. Overall, the ground station equipped with a clock of higher precision not only determines the geopotential difference between two stations at the 1 $m^2/s^2$ level, but also improves the uncertainty of the deviation error. In other words, relativistic PPP time comparison for determining the geopotential difference between two stations is feasible.

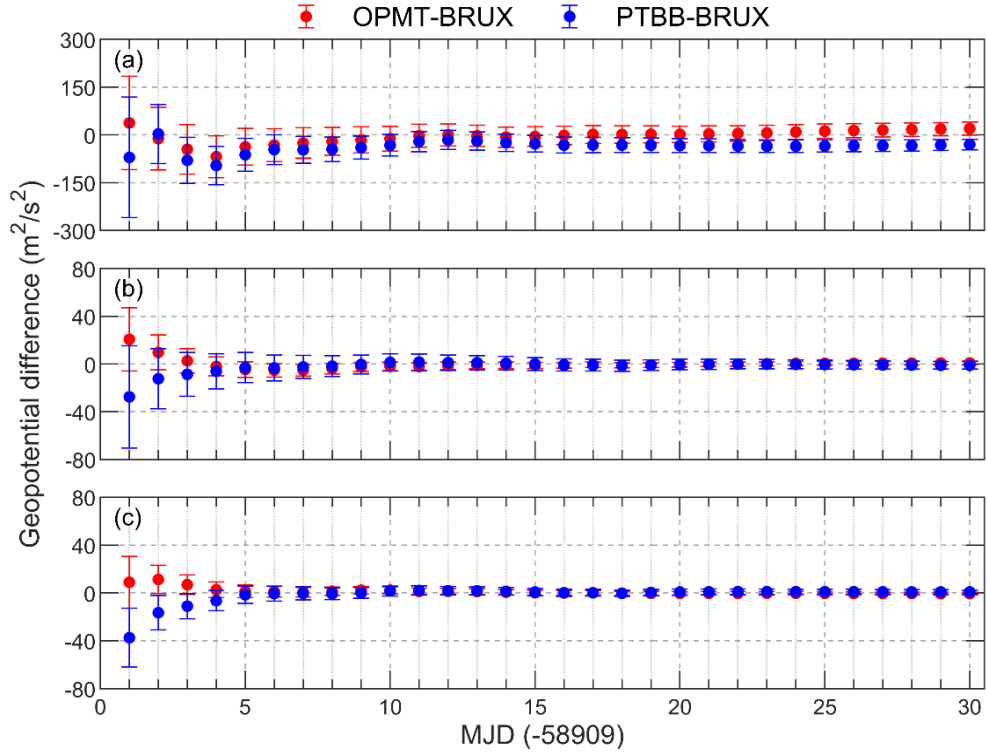

**Fig. 10** Deviation of the determined geopotential difference between two stations from that of the EIGEN-6C4 model, for various lengths of the test period. Subfigures (a)–(c) represent Cases 1–3, respectively. Error bars in (a)-(c) show the 1-σ uncertainty. Each measurement point represents the geopotential difference measured using clock offsets from MJD 58909 to the measurement date

**Table 5** Deviation of the determined geopotential difference between the observed results and the corresponding ones from the EIGEN-6C4 model for OPMT–BRUX during MJD 58909–58938 (unit: $m^2/s^2$)

| Days | Case 1 | Case 2 | Case 3 |
|---|---|---|---|
| 1  | 37.47 ± 146.76 | 20.67 ± 26.49 | 8.87 ± 21.54 |
| 5  | -37.52 ± 57.59 | -4.63 ± 6.56  | 1.31 ± 5.30  |
| 10 | -12.02 ± 38.50 | -1.91 ± 3.60  | 2.24 ± 2.90  |
| 20 | 1.77 ± 25.73   | -0.67 ± 1.97  | -0.39 ± 1.58 |
| 30 | 19.82 ± 20.33  | 0.90 ± 1.39   | -0.64 ± 1.11 |

**Table 6** Deviation of the determined geopotential difference between the observed results and

the corresponding ones from the EIGEN-6C4 model for PTBB–BRUX during MJD 58909–58938 (unit: m$^2$/s$^2$)

| Days | Case 1 | Case 2 | Case 3 |
| --- | --- | --- | --- |
| 1 | -70.81 ± 189.21 | -27.51 ± 42.75 | -37.45 ± 24.59 |
| 5 | -62.29 ± 52.06 | -2.94 ± 12.48 | -1.54 ± 7.12 |
| 10 | -32.44 ± 33.56 | 1.33 ± 7.34 | 1.60 ± 4.17 |
| 20 | -33.54 ± 21.64 | -0.40 ± 4.32 | 0.72 ± 2.45 |
| 30 | -30.13 ± 16.74 | -1.15 ± 3.17 | 0.76 ± 1.79 |

**Discussion and Conclusions**

This study focuses on high-precision time comparison between two arbitrary ground stations for determining the geopotential difference using the relativistic PPP time comparison approach, and comprehensively validating this approach with different receiver clocks. For this purpose, two key requirements need to be satisfied. First, all stations of interests should be equipped with high-precision clocks to maintain a stable frequency standard without artificially adjustments. Second, the clocks used at different stations should be a priori synchronized at one same site. However, it should be acknowledged that the frequency stabilities of the clocks at IGS stations at present are limited to about $1.0 \times 10^{-15}$ at 1-day averaging time. Moreover, a priori synchronization information of two clocks at different IGS stations is not available, and the uncertainty of synchronization affects the accuracy of the experimental results. The key point is that the frequency of the clocks at IGS stations is steered to stay aligned on UTC, so it can certainly not be used to determine the geopotential. Consequently, it is difficult to validate the relativistic PPP time comparison approach at the decimeter level using real GNSS observations at present IGS stations. To solve this problem, we need GNSS observations connected with ultra-high precise atomic clocks which freely run, without being steered to any standard reference. As a prelude, in this study we simulate GNSS observations at ground stations equipped with ultra-high precise atomic clocks without any kinds of adjustments.

We choose three stations and simulate GNSS observations at these stations equipped with atomic clocks with high stabilities to assess the performance of the relativistic PPP time comparison approach over several durations of up to 30 days. The validation is performed following a three-step procedure: (i) simulating GNSS observations, (ii) evaluating the performance of PPP time transfer, and (iii) determining the geopotential difference between two stations. In the first step, we simulated the receiver clock offsets in the three cases based on SDEs, and employed them in the observation simulation. We also analyzed the performance of the clock offsets; the accuracy and uncertainty of clock offsets are the part of errors involved in determining geopotential difference. In the second step, the PPP technique is used to compute the clock offsets and the performance is evaluated. Combining Figs. 6–9 and Table 4, the performance of clock offsets in Case 3 is found to be better than that in Case 1. Regarding frequency stability, results of Cases 2 and 3 improve the long-term stability of OPMT–BRUX as well as PTBB–BRUX. Comparative analyses show that the accuracy of PPP time transfer can be improved by using more accurate clocks, but the improvement is also limited by various noises in observations. Compared to Case 2, the performance of Case 3 is only slightly improved. Based on the performance evaluation of PPP time transfer, it is determined the geopotential difference between two stations during the 30-day period in the last step. With the requirement of maintaining continuous operation of the clocks, deviation errors and uncertainties of time links decrease with increasing experiment time, providing a geopotential difference of high accuracy. We may conclude that the accuracy of this approach is consistent with the stabilities of the clocks used for each case; that is, this approach is feasible for determining the geopotential difference between two stations.

The results of this study show that the relativistic PPP time comparison approach has the potential to achieve decimeter-level accuracy. With the rapid development of GNSS, the accuracy of orbit and clock products as well as the models of various measurement errors will be improved. The determination of geopotential difference between two stations at the centimeter level will be further investigated. The formulation of this study could be also applied to testing GRT.

**Acknowledgments** This research was funded by National Natural Science Foundations of China (grant Nos. 42030105, 41721003, 41804012, 41631072, 41874023, 41974034,

41574007), Natural Science Foundation of Hubei Province of China (grant No. 2019CFB611), China Space Station Project (2020228) and Guangxi Key Laboratory of Spatial Information and Geomatics (grant Nos. 17-259-16-04, 17-259-16-05). We acknowledge the IGS and CODE for providing precise orbit, clock products, and observation data.

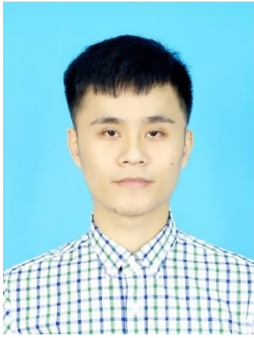 **Chenghui Cai** is currently a Ph.D. candidate at the School of Geodesy and Geomatics, Wuhan University. His current research mainly focuses on relativistic geodesy and GNSS time transfer.

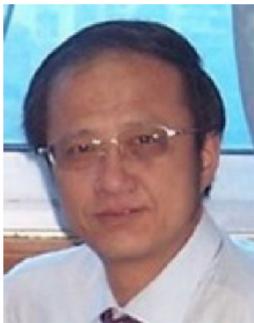 **Wen-Bin Shen** is a head and professor of Department of Geophysics, School of Geodesy and Geomatics, Wuhan University. He received his Ph.D. from Graz Technical University in Austria in 1996. He holds memberships of IAG, EGU, AGU, and IUGG. His research interests focus on relativistic geodesy, Earth rotation and global change, Earth's free oscillation.

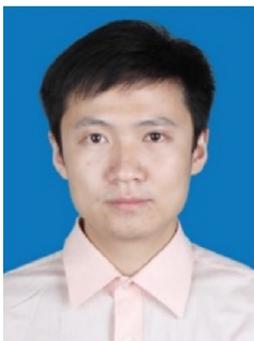 **Ziyu Shen** is currently a lecturer at the School of Resources, Environmental Science and Engineering, Hubei University of Science and Technology, Xianning, China. His current research mainly focuses on relativistic geodesy and satellite frequency transfer technique.

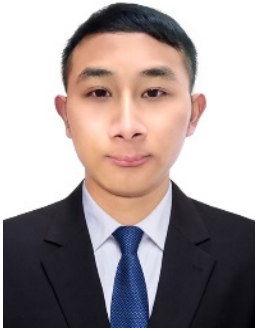**Wei Xu** is currently a Ph. D student at the Wuhan university. He received his master's degree at Anhui University of Science and Technology in 2018. His current research mainly focuses on multi- frequency GNSS data processing and its applications.

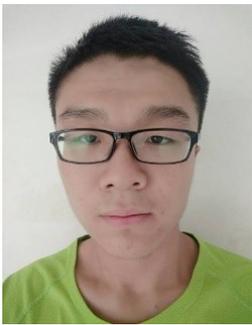**An Ning** is currently a master student at the Wuhan University. His current research mainly focuses on relativistic geodesy based on microwave links between satellites and ground station.